\documentclass[12pt,aaspp4]{aastex}
\input epsf.tex
\begin{document}
\title{Formation of the Terrestrial Planets from a Narrow Annulus}
\author{Brad M. S. Hansen\altaffilmark{1}
}
\altaffiltext{1}{Department of Physics \& Astronomy, and Institute of Geophysics \& Planetary Physics, University of California Los Angeles, Los Angeles, CA 90095; hansen@astro.ucla.edu}


\lefthead{Hansen }
\righthead{Narrow Earth}

\begin{abstract}
We show that the assembly of the Solar System terrestrial planets can be successfully modelled with all of the mass initially confined to a narrow annulus
between 0.7 and 1.0~AU. With this configuration, analogues of Mercury and Mars often form from the collisional evolution of material diffusing
out of the annulus under the scattering of the forming Earth and Venus analogues. The final systems also possess eccentricities and inclinations 
that match the observations, without recourse to dynamical friction from remnant small body populations. Finally, the characteristic
assembly
timescale for Earth analogues is rapid in this model, and consistent with cosmochemical models based on the 
$^{182}$Hf--$^{182}$W isotopes. The agreement between this model and the observations suggests that terrestrial planet systems may also be
formed in `planet traps', as has been proposed recently for the cores of giant planets in our solar system and others.

\end{abstract}

\keywords{planetary systems: formation; solar system: formation; planets and satellites: individual (Earth, Mars, Mercury, Venus)}

\section{Introduction}

The formation of planets is, for the most part, well described by a model in which the available material is
systematically accumulated into fewer but larger bodies, with a variety of physical processes contributing
to the accumulation at different scales (e.g. Safronov 1969; Greenberg et al. 1978;  Lissauer 1987; Wetherill \& Stewart 1993; Kokubo \& Ida 1998; Goldreich et al. 2004).
Traditionally, these processes are held to occur locally, i.e. planets are assumed to grow in radially confined feeding zones, which gradually
increase as the characteristic masses grow. However, the
 extrasolar planet discoveries of the last decade have revealed a number of interesting and surprising
new properties of planetary systems (reviewed recently in Udry \& Santos 2007), including giant planets in short period orbits, planets with highly eccentric orbits
and planets with inflated radii. 
The overall result has been to usher in a new paradigm for planetary system formation, with considerably greater dynamical complexity than the standard solar system formation picture that developed
over the course of the last three centuries. In particular, the notion of large scale radial migration of material calls into question the simple picture of a planetary system
architecture laid down by the initial distribution of mass in the protoplanetary disk. Yet, our own solar system fits the old model relatively well.
 The natural question then arises -- how does our own solar system fit
into this new paradigm? Is ours one of the few planetary systems where planets did, in fact, form in situ, even though
its contemporaries around other stars were engaged in an orgy of migration, scattering, collision and ejection? Alternatively, is it possible to find multiple initial
conditions that produce planetary systems like our own, some of which may be radically different from the standard model, yet potentially applicable to other planetary
systems?

Indeed, in recent years, theories for the early evolution of our own solar system have begun to incorporate a much
more dynamically active description for how the final planetary configurations are produced. 
This has been driven largely by attempts to explain the configuration of the outer planets and their
ambient populations (Fernandez \& Ip 1981; Thommes, Duncan \& Levison 1999; Tsiganis et al. 2005; Gomes et al. 2005).
 For the purposes of what is to follow, perhaps the most telling feature of these models is the
fact that the initial conditions for the dynamical evolution depart significantly from the traditional 'Minimum Mass Solar Nebula' surface density profile (Weidenschilling 1977a; Hayashi 1981), in that the outer planets are hypothesized to have begun their evolution
much closer to Jupiter than they are found today, implying a somewhat more localised concentration of material. Whether this configuration represents the effect of some chemical
transition such as the 'ice line' or a dynamical trapping mechanism is still a subject of discussion. However, as an initial hypothesis it allows for a dynamical model that explains a wide variety of observed properties.

In this paper, we demonstrate that the configuration of the inner solar system can also  be explained by starting
with a considerably more localised initial condition. Furthermore, this model can explain 
several features of the observations that have proven difficult to explain with more traditional starting conditions. In essence, we postulate that {\em all} the rocky material that eventually came to comprise the terrestrial planets
was initially restricted to a narrow annulus. We show that (\S~\ref{Annulus}), starting with this configuration, 
the accumulation process not only results in the formation of Earth and Venus-like objects, but also the formation
of Mercury and Mars analogues, both with the appropriate location and masses. This is in spite of the
fact that the initial conditions have no material at the present locations of these planets.
 In \S~\ref{Discuss} we review how the
results compare with dynamical and cosmochemical constraints from the solar system.

\section{Terrestrial planet Formation in a narrow annulus}
\label{Annulus}

Current thinking divides the formation of rocky bodies into several epochs, beginning with the
accumulation of dust particles into planetesimals, followed by the accumulation of planetesimals into planetary
embryos, which eventually leads into the epoch where the planetary embryos collisionally evolve to leave behind
the planetary system we observe today. This last stage, in which the mass is distributed primarily amongst a few
tens to hundreds of remnant bodies, is the one of interest to us. In the earlier stages, the dynamics of the accumulation
is essentially local, and it is only when the masses of the largest bodies become substantial enough to affect the dynamics 
of similar masses in neighbouring annuli that the global properties of the protoplanetary system become important.

Over the past decade, the evolution of computer processors and algorithms has made these latter stages amenable to
direct numerical simulation. Such studies (Chambers \& Wetherill 1998; Agnor et al. 1999; Chambers 2001; Raymond
et al. 2004; O'Brien, Morbidelli \& Levison 2006) have demonstrated a broad agreement with the observations, in
the sense that they tend to produce planetary systems which have approximately the right number of planets, with
approximately the right mass and in approximately the right location. However, some quantitative failures of the
generic model remain. Although simulated analogues of Earth and Venus emerge naturally, the simulations have
trouble reproducing the observed characteristics of Mars and Mercury. In particular, while planets are often found
in the right location, they are almost always too massive. Furthermore, most of the simulations produce planetary
systems that are too dynamically `hot', in the sense that the inclinations and eccentricities of the modelled planets are too large compared to those observed.
 A potential solution to this latter problem is the addition of a dynamically significant population of small bodies, which provide a source of dynamical friction that can damp the eccentricities and inclinations (O'Brien et al. 2006). The late-time existence of a dynamically important population of small bodies is not guaranteed, however, and another potential solution is excitation and merger driven by
secular resonance sweeping from the giant planets (Thommes, Nagasawa \& Lin 2008) as they migrate in a dissipating gas disk.
 A final question mark is the time it takes the simulated systems to reach their final configuration -- in many scenarios it may be somewhat too long compared to various cosmochemical measures of the Earth's formation
timescale (we will return to this point in \S~\ref{Discuss}).

Recently, we chose to study the formation of the extrasolar terrestrial planet system orbiting the pulsar
B1257+12 (Wolszczan \& Frail 1992; Konacki \& Wolszczan 2003), with the goal of possibly gaining insights into
the formation of terrestrial planets with a completely independent system (Currie \& Hansen 2007;
Hansen, Shih \& Currie 2008). The results of that study suggest that the best model for the formation of the
pulsar planets was formation from a narrow ring of solid material, deposited as an expanding disk of post-supernova debris cooled and deposited heavy elements at the outer edge. Although the conditions are very different between this
system and the solar system protoplanetary disk, the similarity in the final configurations  motivated us to consider the question of whether the
hypothesis of a narrow ring of material as an initial condition would also suffice to explain the terrestrial
planets of the solar system.

Let us consider then, 
 the following question of principle -- if we start with $2 M_{\oplus}$ in small bodies, confined to a narrow annulus between 0.7--1~AU, what is the likely outcome? We have used the
planetary dynamics code {\em Mercury6} (Chambers 1999) to simulate this scenario. For
our default model, we assume a uniform surface density in solids, between 0.7--1~AU, broken up between 400
equal mass bodies, each of mass 0.005$M_{\oplus}$\footnote{Most traditional simulations start with a mass spectrum, estimated from theories of oligarchic growth. For the
high surface densities considered here, the resulting isolation masses would already be of planetary scale. Thus, we are effectively starting our simulation before the
end of the oligarchic accumulation process.}, for a total mass of 2$M_{\oplus}$. The semi-major axes are spread randomly within the annulus\footnote{We also experimented with equal spacing and found no identifiable difference in behaviour.} and all orbits are assumed to
be initially circular and with a small ($\sim 0.5^{\circ}$) dispersion in inclination.
The timestep is taken
to be 4 days, and the stepsize accuracy constraint is $\epsilon = 10^{-11}$. The systems are run for 10$^9$ years,
to allow the completion of the accumulation into planets.
 For our baseline set of models,
we also include the presence of Jupiter at 5.2~AU and eccentricity=0.05, as the resulting gravitational perturbations can have a quantitative
effect on the outcome of terrestrial planet accumulation (see discussions in Chambers 2001; O'Brien et al. 2006).

\section{Results}

In the case of the outer planets, the masses are large enough that the scattering of planetesimals back and forth
promotes an effective repulsive force, resulting in the outward migration of Uranus and Neptune. However, in the
inner solar system, the potential well is deep enough and the masses small enough that it
 is not sufficient to drive significant
radial migration, and we expect most of the material to accumulate into a handful of earth-size planets, roughly in situ. The results of 23 realisations of this scenario are shown in Table~\ref{CompTab} where indeed,
this does occur and, for such a narrow annulus, the conservation of energy and
angular momentum results in a tendency  to produce two roughly equal mass
planets, located near the inner and outer edges of the original annulus. Thus, the formation of Earth and Venus
analogues is hardly surprising in the context of this model. What is perhaps more surprising is the fact that
the simulations frequently produce analogues of Mercury and Mars, both in mass and location. This is something
that previous simulations have not successfully reproduced with any regularity (e.g. Chambers 2001). Figure~\ref{am} shows the accumulated
sample of all planets surviving in the simulations after $10^9$ years. As expected, the distribution in mass and semi-major
axis peaks at the edges of the original annulus, yielding the Earth and Venus analogues. However, we see that several
Mercury analogues and an abundance of Mars analogues also remain. The group of Mercury analogues (defined here
formally as $a<0.65$~AU and $M<0.2 M_{\oplus}$) are clearly distinct
from the larger bodies, and have an overall abundance in the simulations of 3/23=0.13. Mars analogues ($a>1.3$AU, $M<0.2 M_{\oplus}$)
have an abundance 21/23=0.91. If we restrict the latter to also have $M > 0.02 M_{\oplus}$ (in order to exclude unaccreted embryos),
the abundance is 13/23=0.57.

Figure~\ref{Planet2} shows the results for six of these simulations, after $10^9$~years, with the size of the point
scaling with the planetary diameter (i.e. as the cube root of mass) and the horizontal
error bar indicating the radial motion due to the eccentricity. We have chosen specific examples to illustrate the
nature of the outcomes. The top two simulations are ones that produce a Mercury analogue, and possess several features of
the observed system. Intermediate cases also occur, in which we are left with three bodies with mass $>0.2 M_{\oplus}$ and
somewhat more closely packed, sometimes with a Mars analogue as well.


A more quantitative comparison can be made using some of the statistics collected by Chambers (2001). Figure~\ref{Splot3} shows the distribution of three statistics; $S_d$, the normalised angular momentum deficit (a measure of the average deviation from
circular, coplanar orbits in the system); $S_s$, a measure of how closely packed the planets are in terms of their own
Hill radii,  and $S_c$, a concentration statistic (measuring how narrowly concentrated the
overall mass distribution is in radius). In each histogram the vertical dotted line indicates the value of the observed terrestrial planet system.
We see that the range in each statistic obtained from the simulations nicely encompasses the observed system. As noted above, traditionally simulations have had difficulties matching the
observations to the same degree, although recent models that include a population of small
bodies do resolve the largest discrepancies , by virtue of the enhanced dynamical friction (e.g. O'Brien et al. 2006).

The fact that the observed masses of Mercury and Mars are not larger can be considered an argument for the
localised annulus initial condition. To illustrate this, we performed two simulations in which we modified our initial
conditions so that 300 of the bodies were in the original annulus from 0.7--1~AU, but 50 bodies were distributed in the regions 0.5--0.7~AU and 1--1.2~AU respectively. Thus, there was 0.25$M_{\oplus}$ available in each of these neighbouring annuli. In each case, a planet formed between 0.5--0.6~AU, but the masses were 0.43$M_{\oplus}$ and 0.65$M_{\oplus}$ respectively. Thus, $\sim 40-60\%$ of the final mass of the innermost planet was ``pulled out'' of the central annulus, because the initial seed material collides with and accretes bodies scattered out of the central annulus. In both cases, a Mars analogue formed. In one case it was of appropriate mass ($0.05 M_{\oplus}$) and in
another it was too large ($0.41 M_{\oplus}$). Both cases also contained a remnant, unaccreted embryo on an orbit outside the Mars analogue.
 This increases the probability of collisions and allows more mass to be accumulated and decoupled from the central concentration, leading to larger masses in the Mercury and Mars regions. In essence, the fact that Mercury and Mars can be explained entirely by scattered material argues for a very narrow initial distribution of mass, particularly at the inner edge.

\subsection{Origin of Mercury}

The path to the formation of Mercury as a remnant of an inwardly diffusing tail of material is illustrated in Figure~\ref{Merc3}. The proto-Mercury initially grows to about half
it's final size in the original annulus location but then slowly diffuses inwards, due to dynamical friction with the inner edge of the diffusion tail. After 10~Myr,
the situation is as shown in 
Figure~\ref{Merc3}. The solid lines show the orbits of the proto-Mercury and proto-Venus. The two orbits do not come close to crossing but are still dynamically coupled
by a variety of smaller bodies whose more eccentric orbits cross those of the two larger bodies. Eventually, these intermediaries are removed by collisions with both
the Mercury and Venus analogues (and a small fraction via ejections). With the eventual clearing of this intermediate population, the system settles down to its final
configuration. 

Sometimes the decoupling process can be more sudden. In Figure~\ref{Merc1}, we show the evolution of a system in which the Mercury continues to scatter off a larger outer
body for $\sim 50$~Myr. However, in this case, the system has two bodies, at $\sim 0.6$~AU and 0.8~AU, which eventually collide to form the final Venus analogue. The product
of this late collision has a semi-major axis at $\sim 0.7$~AU, which means the Mercury is now decoupled from the outer bodies. Similarly, Figure~\ref{Merc4} shows a case
in which the proto-Mercury suffers two major collisions in short succession -- the first of which displaces the body to an interior orbit that still crosses that of the
proto-Venus, and the second of which reduces the eccentricity to the point that the Mercury is thereafter dynamically decoupled. It does undergo further collisions later,
but they do not bring it back into contact with the main mass reservoir.
 So the dynamical decoupling can occur either by the erosion of an intermediate population of small bodies, or a rapid adjustment of the orbit of one of the principals.

The prominent role of collisions in the formation of the Mercury analogues also accords well with one of the scenarios to explain Mercury's anomalously high density (Urey 1951; Harder \& Schubert 2001). 
The several large late impacts evident in Figure~\ref{Merc4} provide a mechanism for blowing off a Silicate-rich mantle to leave behind an iron-rich core, as has been discussed by several authors (Smith 1979; Wetherill 1986; Cameron et al. 1988; Benz et al. 2007).

\subsection{Origin of Mars}

Mars analogues are much more common, occuring in most of the simulations. However, there is some range in the properties from simulation to simulation. As can be seen
in Figure~\ref{am}, there is a significant population whose mass and location is an excellent match to the observed Mars. However, in roughly 1/3 of the cases, the remaining surviving body is an unaccreted embryo and thus significantly less massive.

Figure~\ref{Mars1} illustrates the origin of one of these bodies. As in the case of Figure~\ref{Merc3}, a body at the outer edge of the diffusion tail (now outwards rather
than inwards) gets circularised to the point that it no longer crosses the orbit of the next closest large body (the proto-Earth). However, it is still dynamically coupled
via it's interactions with intermediary bodies. Over time, these bodies are removed (by accretion, ejection and occasionally collisions with the Sun). If the body
retains a small eccentricity during this clearing-out stage, it will survive.

The route to a more traditional Mars is shown in Figure~\ref{Mars2}. The proto-Mars is scattered out of the main planetary region relatively early, but remains
strongly coupled via its interactions with several other scattered bodies. The accretional growth occurs quite rapidly (within 10 Myr) but the body remains strongly
coupled for timescales $> 100$~Myr because there are still several bodies on crossing orbits at these late times. This is a generic difference between Mercury and
Mars histories, as the diffusion tail outwards is larger and longer-lasting than that which extends inwards (even though it is limited by the gravitational influence
of Jupiter). Figure~\ref{Mars2} shows the orbital evolution of the proto-Mars and proto-Earth, with the removal events of the last scattered bodies indicated. Most
of these are accreted by the Earth (although the penultimate event is an ejection). 

The distribution of eccentricity with semi-major axis for all surviving bodies is shown in Figure~\ref{ae}. Both the Mercury and Mars analogues fall nicely within
the trend of increasing eccentricity as one moves away from the central annulus (as expected based on the proposed scenario). The simulated
points in Figure~\ref{ae} are divided into two classes, those with mass $> 0.02 M_{\oplus}$ and those with $< 0.02 M_{\oplus}$. This is because
the objects with the highest eccentricities are largely unaccreted embryos, and substantially smaller than Mercury or Mars. 
The eccentricities
of Venus and Earth lie at the low end of the simulated distribution, but the fact that the simulations produce any eccentricities that low is
encouraging. The distribution of inclination (with respect to the most massive body in each simulation) with semi-major axis is shown in Figure~\ref{ai}. Mercury is found plausibly at the edge of the inward tail, albeit with small statistics, while Mars appears to fall at the low end of the distribution, but well
within the spread (5/12 of the simulated bodies with $M >0.02 M_{\oplus}$ and $a>1.3$AU have inclinations less than twice that of Mars). 


\subsection{Small Bodies, Dynamical Friction and the Asteroid Belt}

It is also interesting that we do not need to incorporate any extra population of small bodies to reduce the eccentricities of the final objects (see Figure~\ref{Splot3}).
 In part this may be because bodies growing from a more extended mass distribution have to reach higher eccentricities to cross the orbits of, and accrete, the material at the edges of the distribution. It is also partially true that our simulations naturally contain elements of the small body damping discussed by O'Brien et al, in the sense that we start with a large number of embryos of small mass.
However, we do not  make any assumptions about the relative mass in small versus large bodies (and so do not need to invoke any specific model for (re)generating the small body population from collisional debris), 
because we do not start with an oligarchic spectrum\footnote{By concentrating the mass in such a narrow range, a traditional estimate for an oligarchic isolation
mass would yield planet-size masses at the start.} of masses and
therefore the size of the respective small and large body mass reservoirs emerge self-consistently from the simulation.

We note that, although we have success in producing the planets of the inner solar system, we do not produce any long-lasting bodies with $a>2$AU, i.e. nothing that
might indicate the production of an asteroid belt. This is no doubt due, in  part, to the barrier from the Jovian 4:1 mean motion resonance (if Saturn were included,
the $\nu_6$ secular resonance would strengthen this barrier). Another possible
contributing factor is that we simply do not have the mass resolution to properly probe the dynamics of that small a fraction of the mass reservoir. Indeed, Bottke et al. (2006) find that iron-rich meteorites in the inner asteroid belt can indeed have originated in the terrestrial planet zone, but that the trapping efficiency is $<10^{-3}$ from a location in our annulus. As such, our simulations would not produce a significant population. Thus, while
 it is
traditionally held that the region of the asteroid belt initially contained much more mass and was gradually depleted by gravitational torques, it is possible that
this region simply represents the far end of a diffusion tail from the terrestrial annulus (although the volatile rich outer belt suggests that at least some of
the material must have originated further out). An alternative hypothesis is that the asteroid belt represents the residue
from an incomplete clearing by whichever process served to accumulate the terrestrial planet material in the first place. 

\subsection{The Role of Jupiter}

The above simulations include the perturbations from Jupiter located at 5.2~AU, with an eccentricity e=0.05. These serve to limit the outward diffusing tail of planetesimals,
and thereby speed up the accumulation process. As an illustration, in Figure~\ref{NoJup}, we show a snapshot of the outer terrestrial zone at 100~Myr in a simulation with
no giant planet perturbations. One can see that a couple of bodies have orbits that take them outside of 2~AU and that there are more bodies present than at a similar
stage in the other simulations. Nevertheless, the end result of this system after $10^9$ years is not very different. Only one Mars analogue remains at
late times and it is located at 1.9~AU, a little
further out than normal but still interior to the location of the asteroid belt.

The calculations above were performed with a Jupiter in the system from the beginning (for reasons of simplicity). However, it takes a finite amount of time to accrete
a core of $\sim 10 M_{\oplus}$ at 5~AU, so we have also calculated 15 additional simulations in which Jupiter is only introduced after 5~Myr. The instantaneous insertion of additional perturbations is a rather crude approximation, but it turns out to have little influence. Figure~\ref{am2} shows the equivalent of Figure~\ref{am} for this new set of simulations.
We see that the shape of the final distribution of objects is very similar, and with similar proportions of Mercury and Mars analogues. The one potential difference is
a larger number of bodies intermediate between Mercury and Venus. Overall, this is perhaps not too surprising as the effect of Jupiter is greatest on the orbital evolution
of Mars analogues, and those can take $\sim 100 Myr$ to complete their evolution.

If we combine the numbers from the two simulations, then the incidence of Mercury analogues is 6/38 = 16\% and Mars analogues is 32/38=84\% (21/38=55\% if we include
a lower mass cutoff $\geq 0.02 M_{\oplus}$. We can increase the incidence
of Mercury analogues to 10/38=26\% if we increase the mass limit to $0.3 M_{\oplus}$. Regardless of the specific value, we recover objects similar to Mars and Mercury at
an interesting rate.

\subsection{Timescales and Cosmochemical Constraints}

Cosmochemical studies of radionuclides can provide valuable constraints on the timescale over which planets accumulate their mass, although the inferences are somewhat dependant on the details of the isotopes and methods used
(e.g. Halliday 2004; Jacobsen 2005;  Touboul et al. 2007). For our purposes, we demonstrate the
implications of our scenario by comparing to the results based on the  $^{182}$Hf--$^{182}$W isotope system, reviewed
in 
 Jacobsen (2005).
Jacobsen infers that the bulk of Earth's core (formally defined as 63\% of the final mass) must have accumulated within $11 \pm 1$ years
and that the moon formed from a major impact at the cessation of Earth's accretion after 30 Myr. This was found to be roughly a factor of two shorter than expected from
N-body simulations of the traditional accretion scenario.

Cosmochemical studies are capable of inferring both the timescale for the accumulation of the bulk of the Earth's material
and also the timing of the last collision with a large body, which is believed to give rise to the formation of the moon.
Therefore, in Figure~\ref{Tx} we show the cumulative distribution of three different timescales as drawn from the
simulations shown in Figure~\ref{am}. For each simulation, we consider the largest of the final remaining planets, and
record the timescale to accumulate 63\% and 90\% of the final mass, and the time of the last impact with a body of mass
$>0.02 M_{\oplus}$ (a generous interpretation of the criterion that the moon-forming impact requires a Mars-size body). 
The distributions show a rather broad range, with median values of 5 Myr, 17 Myr and 45 Myr respectively, but with broad
ranges. Figure~\ref{Three} shows three examples of the largest body histories, from which these distributions were derived.
They were chosen to represent the range of 63\% accumulation times, and show a variety of histories, with roughly exponential
accumulation in one case, a very large impact at early times in another, and a large impact at late times in the third case.
This variety means that the simulations can accomodate a variety of proposed scenarios. As an example, Jacobsen et al. (2009)
present two scenarios consistent with the cosmochemical evidence. One requires that 63\% of the Earth form within 11 Myr and
the Earth-Moon system formed at 32~Myr. The other requires that 90\% of the earth formed within 6~Myr and that the Earth-Moon
system formed late ($\sim 100$~Myr). The distributions in Figure~\ref{Tx} favour the first scenario slightly, but can accomodate the second scenario at the 2$\sigma$ level. Another prominent cosmochemical result is that by Touboul et al. (2007), which claims that the Moon-forming impact must have occurred within 50~Myr of the Earth's formation -- quite consistent with our median last impact time of 45~Myr.
One final point to note is that some of the late-time impactors in these simulations are considerably larger than the Mars-mass impactors traditionally
used in simulations of the moon-forming impact (e.g. Canup 2004).


Figure~\ref{TM} shows the accretional history of the system shown in Figure~\ref{Mars2}. We see that both the Mars and Earth analogues accumulate the bulk of their mass
within 10~Myr, but the Earth has an accretional tail that extends out to somewhat longer times. This is simply a reflection of the larger mass and hence larger cross-section
for accretion -- so that collisions with Earth are much more important in the final clearing stage than collisions with Mars. The rapid accumulation of Mars is also
nicely consistent with the cosmochemical constraints (Nimmo \& Kleine 2007 review the cosmochemical evidence and suggest ranges from 1-10~Myr). 

Another important line of cosmochemical evidence is the identification of isotopic differences in the compositions
of the various planets (e.g. Lodders 2000). The interpretation of these differences in terms of the primordial
mass reservoir from which the planets accrete is not easily translated to the current scenario, since most analyses
make an initial assumption that the various families of meteorite represent the original planetary mass reservoir.
Our model produces no analogue to the asteroid belt (even the remnant unaccreted embryos that sometimes survive
lie at $a<2$~AU) and so it would require additional assumptions, beyond the scope of this idealised model, to make
a proper connection with those analyses. However, it is potentially possible to get compositional differences if
the gradient across the narrow annulus is large enough. Figure~\ref{Grad} shows the cumulative distribution of
initial positions of the embryos that that end up in the Mars analogues of two of our simulations. We see that,
in one case, the material that goes into the Mars is well-mixed, and representative of the overall distribution,
while the other case is composed preferentially from material drawn from the outer part of the annulus. Thus, it
is possible to get some difference, although not in any predictable fashion (essentially because the Mars body is a
scattered remnant and thus subject to the vagaries of the chaotic dynamics of the system).

\section{Discussion}
\label{Discuss}

The model discussed in this paper differs from traditional accumulation simulations (e.g. Chambers 2001)
only in that we take seriously the notion that the initial conditions might be radically different from
those of traditional simulations. However, this assumption does provide an explanation for several previously
puzzling features, such as the masses of Mercury and Mars, the low eccentricities of the final planets,
and the rapid accumulation times inferred in cosmochemical studies. This model also finds some
level of support in other studies which have touched on one or more related aspects (albeit sometimes
accidentally!). The idea that the Mars region might be depleted in initial mass due to chemical depletion
in a cooling nebula has been proposed before (Chambers \& Cassen 2002; see also Jin et al. 2008), although the
specific realisations of that scenario did not match the observations well. Similarly, Morishima et al. (2008)
start with a narrower annulus than usual, although primarily for computational convenience. They do produce 
occasional Mars-like bodies, but do not produce Mercurys, because their annulus extends too far inwards.
The localisation of the initial conditions in previous models was simply not extreme enough to properly
characterise the diffusive tails that lead to the natural mass and location of Mars and Mercury.

Of course, the initial conditions used here are quite extreme, in the sense that {\em all} the mass available
to form terrestrial planets was restricted to a narrow annulus. The goal is to
demonstrate the consequences of this choice of initial condition and that this allows us to
match the observations of solar system planets, in a variety of measures. Test simulations with broader distributions
suggest that the narrowness of the concentration is an important component of the model's success, although it is 
possible that the mass of Mercury is more sensitive to this than the mass of Mars (so that a sharp inner edge may
be more important than a sharp outer edge). The degree to which a true physical model can realise these conditions
is unclear, but the fact that such an initial condition could potentially resolve several long-standing problems
makes their consideration worthwhile. 
To that end, there are several possible avenues of investigation.

The notion that planetesimals form by gravitational instability (Goldreich \& Ward 1973) has recently
received a resurgence. The original theory was hampered by concerns that turbulent stirring (Weidenschilling 1977b)
would prevent the particle layer from achieving sufficiently high densities to become unstable.
In recent years, Youdin \& Shu (2002) have advocated that 
gravitational instability can overcome the mixing if sufficient overdensities in solid material are
created. This can potentially be realised if radial migration of solid material generates particle pile-ups
(Haghighipour \& Boss 2003; Youdin \& Chiang 2004; Johansen et al. 2007) at preferred locations in the disk. Such pile-ups may eventually produce a preferred
annulus for accumulation of planetesimals (Youdin 2005), as required by this scenario.

Alternatively, if planetesimals are subject to migration as the result of hydrodynamic drag or density wave
dissipation, the radial motion can be quite rapid (Goldreich \& Tremaine 1980; Ward 1997). This inward migration
of solid material can remove nearly all the solid material from the disk unless the migration is halted somehow.
It has been suggested (e.g. Masset et al. 2006) that local changes in the density profile, either due to changes in disk structure or
true evacuation of inner gaseous holes, can reduce or reverse the effects of this migration, which would also
lead to a pile-up of solid material. The temperature structure may also halt migration for planetary mass bodies
(Pardekooper \& Mellema 2006), although this may not work at the masses of embryos. The properties of migration
may also change if the character of the disk changes with location. The interface
between active and dead zones in a protoplanetary disk may mark a transition from laminar behaviour to turbulent behaviour,
which may also affect the manner in which migration occurs (Rice \& Armitage 2003; Johnson, Goodman \& Menou 2006), 
and density gradients at the interface may also contribute (Kretke \& Lin 2007). Thus, there are several potential mechanisms
for accumulating radially migrating material at specific radii.

While it is encouraging that the hypothesis of a narrow initial annulus helps to explain several of the nagging problems
of solar system terrestrial planet formation, the most interesting implications are what similar configurations might lead
to in extrasolar planetary systems. This will become particularly important in the near future, as both radial velocity and
transit searches push further down into the earth-mass regime. However, assessing these implications is, at present, problematic
because the greatest observational sensitivity to low mass planets is around lower mass stars, and without a specific physical mechanism
for the location of the annulus, it is difficult to forecast the expected location around stars of different types in general
terms -- it will depend on the particular model.

\section{Conclusion}

We have shown that the observed system of terrestrial planets are well reproduced by gravitational assembly of
a system of planetary embryos initially confined to a narrow annulus between 0.7 and 1~AU. With this initial
configuration, the planets form rapidly (the bulk accreting within 10~Myr) -- consistent with inferences from
$^{182}$Hf-$^{182}$W chronometry -- and Mercury and Mars analogues form from bodies that are scattered out of
the original annulus and dynamically decoupled by subsequent interactions. The results of these simulations
agree well with various statistical measures of the solar system configuration and reproduce the masses and
location of the planets with more fidelity than similar calculations that start with more traditional power
law density profiles.

\acknowledgements BH acknowledges support by the Sloan Foundation, the 
NASA Astrobiology Society and the Space Interferometry Mission. In particular, the computing
cluster purchased with Sloan Funds made the calculations presented in this paper
considerably less painful than they would otherwise have been. He also acknowledges informative
comments from Phil Armitage, Eric Gaidos, Scott Tremaine, Andrew Youdin and Ed Young, as
well as the participants of the Aspen Summer Workshop on the Formation and Habitability of 
SuperEarths.

\newpage

\newpage
\plotone{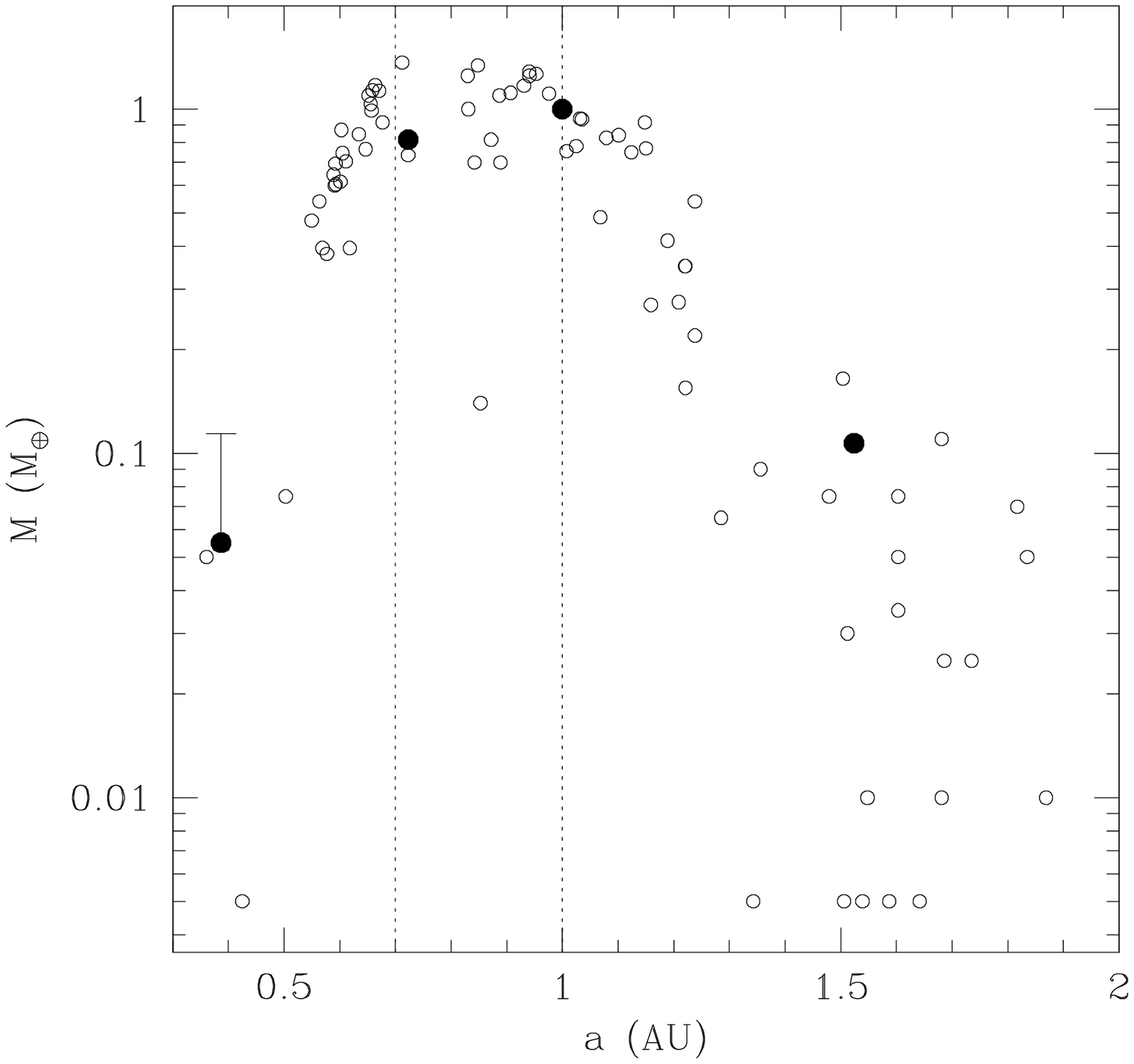}
\figcaption[Example]{The open circles represent the distribution of bodies remaining in our simulations, while the 
solid points are the four observed solar system planets. The vertical error bar for Mercury indicates the possible
original larger mass if it originally had the same iron content as the other terrestrial planets. The vertical dotted
lines indicate the edges of the original annulus.
 \label{am}}
\newpage

\plotone{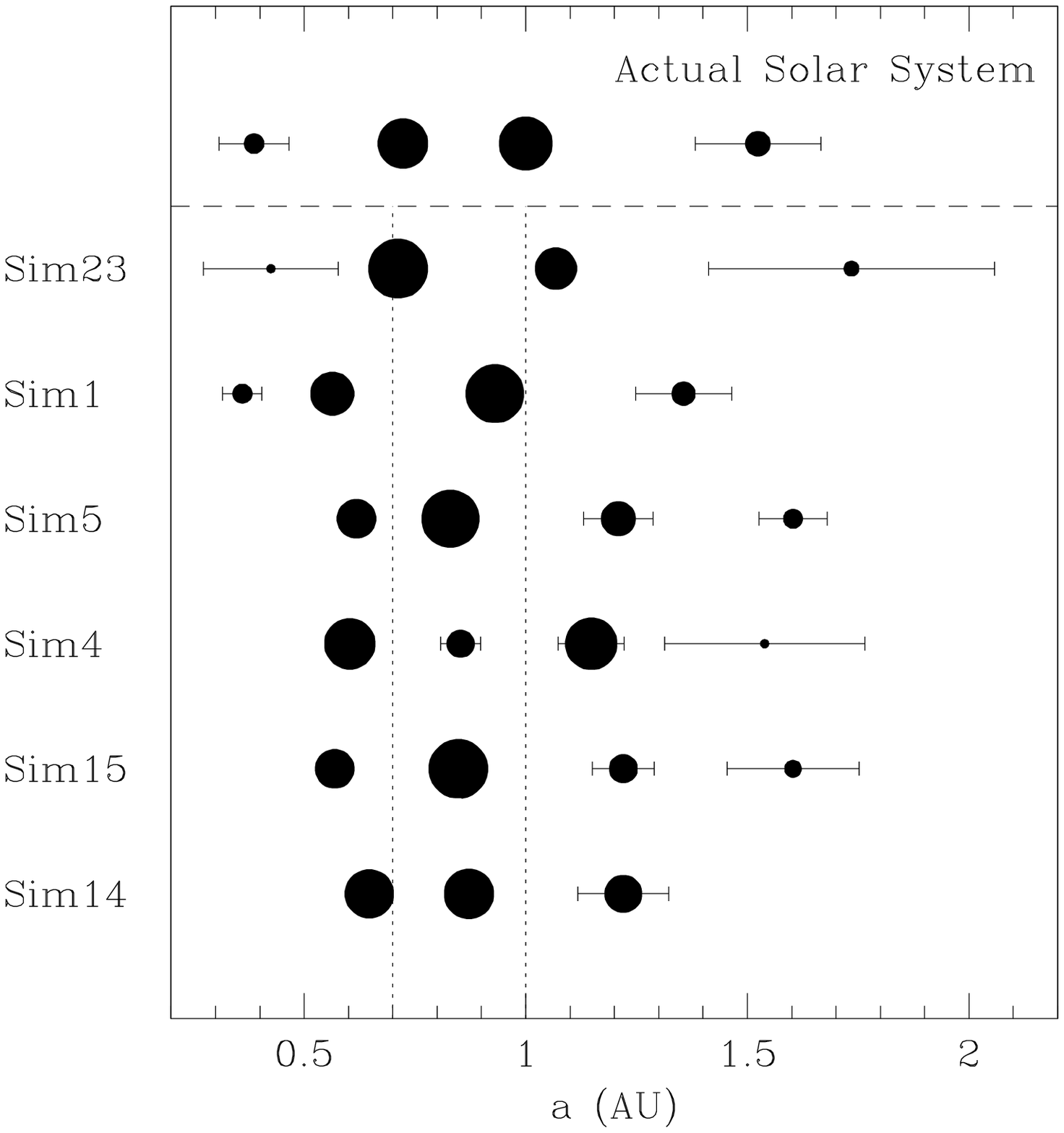}
\figcaption[Planet2]{The top system is the observed terrestrial planets, Mercury, Venus, Earth and Mars. Below
that are six realisations of a simulation which begins with 2 earth masses of material spread uniformly between
0.7 and 1~AU (as indicated by the vertical dotted lines). The size of the plotted points scales as the cube root
of the planet mass, i.e. approximately with the linear dimensions. The horizontal error bars indicate the radial
excursions that result from the planetary eccentricity.
 We see that Earth and Venus analogues form naturally around the location of the annulus, while
Mercury and Mars analogues are often produced by remnant bodies that are scattered out of the forming region and
eventually become dynamically decoupled.
 \label{Planet2}}
\newpage

\plotone{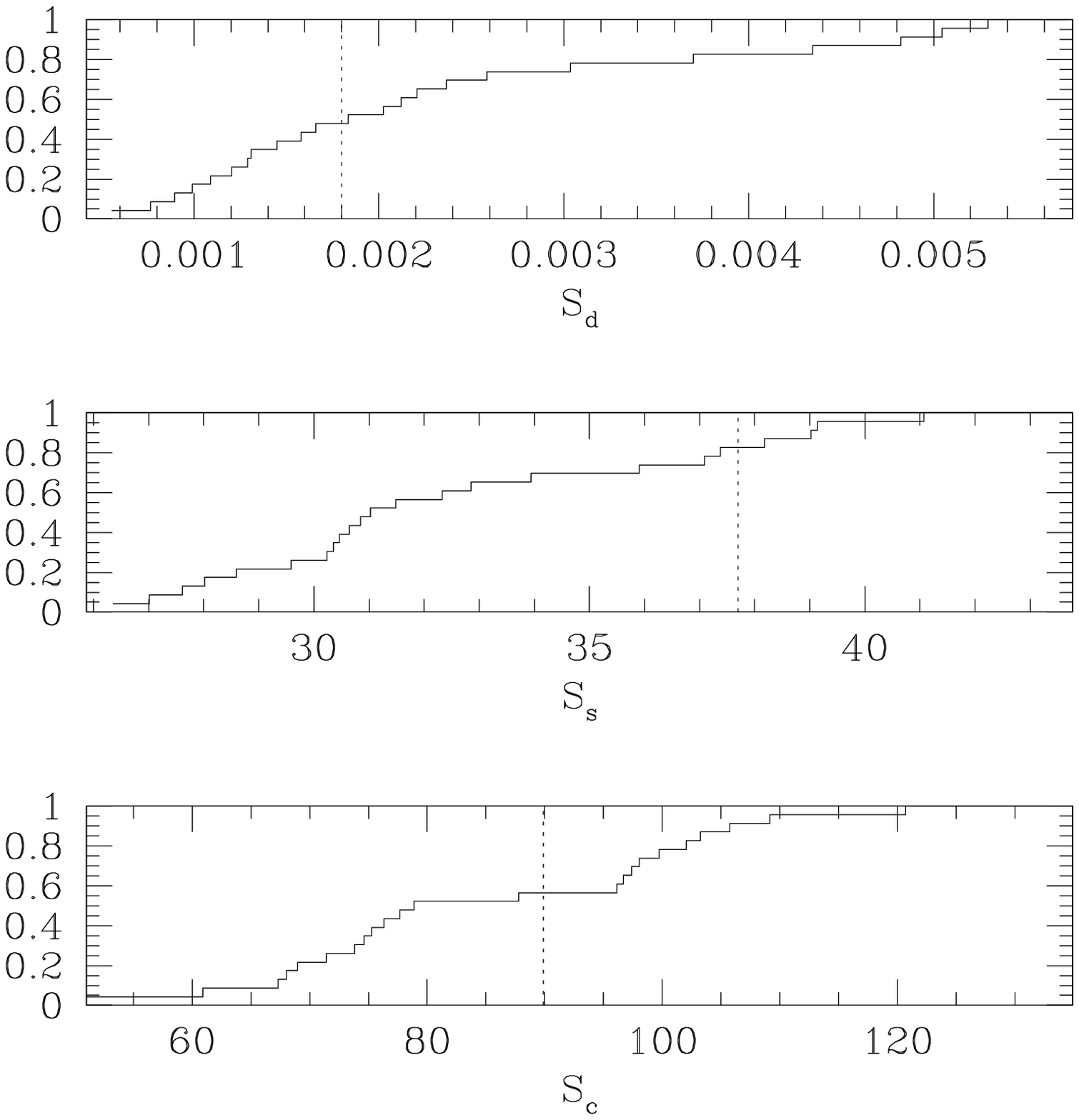}
\figcaption[Splot3]{Each panel shows the cumulative probability distribution of the statistics discussed in the text,
as evaluated for the 23 simulations listed in Table~\ref{CompTab}. $S_d$ refers to the normalised angular momentum deficit, 
$S_s$ is an orbital spacing statistic, and $S_c$ is a mass-concentration statistic.
 The vertical dotted lines indicate the same statistic
calculated for the observed Mercury-Venus-Earth-Mars system. We see that the solar system falls well within the range in
each case.
 \label{Splot3}}
\newpage

\plotone{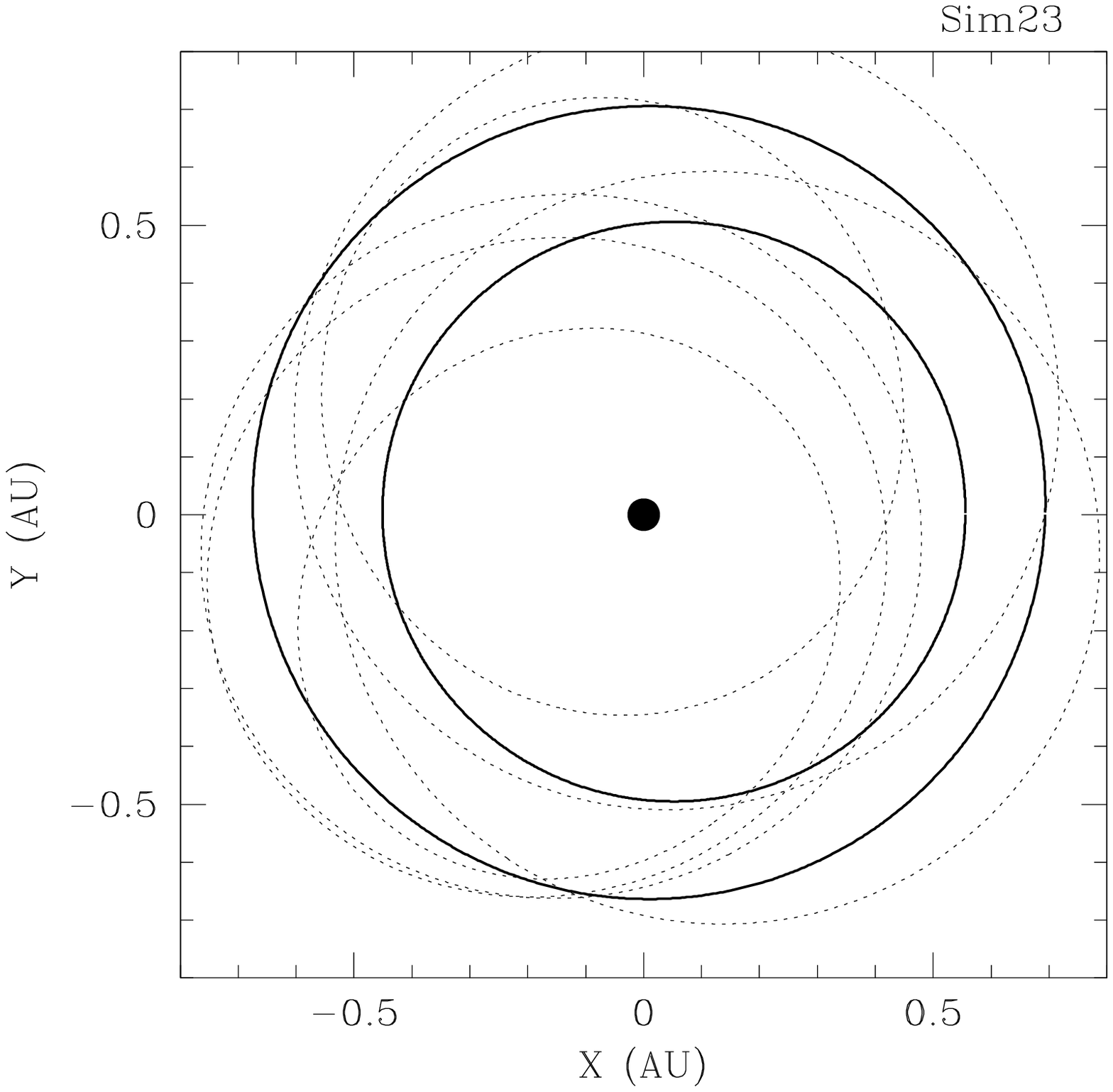}
\figcaption[Merc3]{The two solid lines show the orbits for the proto-Mercury and proto-Venus in one of the simulations,
after 10 Myr of accumulation. The dotted lines show the orbits of smaller bodies whose semi-major axis lies between these
two. (There are many other bodies which lie exterior to the proto-Venus, whose orbits are omitted for the purposes of
clarity). These bodies have markedly more eccentric orbits and the scattering of these back and forth serve to dynamically
couple the larger bodies. As the smaller bodies are slowly removed, primarily by accretion, the larger bodies become dynamically 
decoupled and settle down to the final configuration.
\label{Merc3}}

\newpage

\plotone{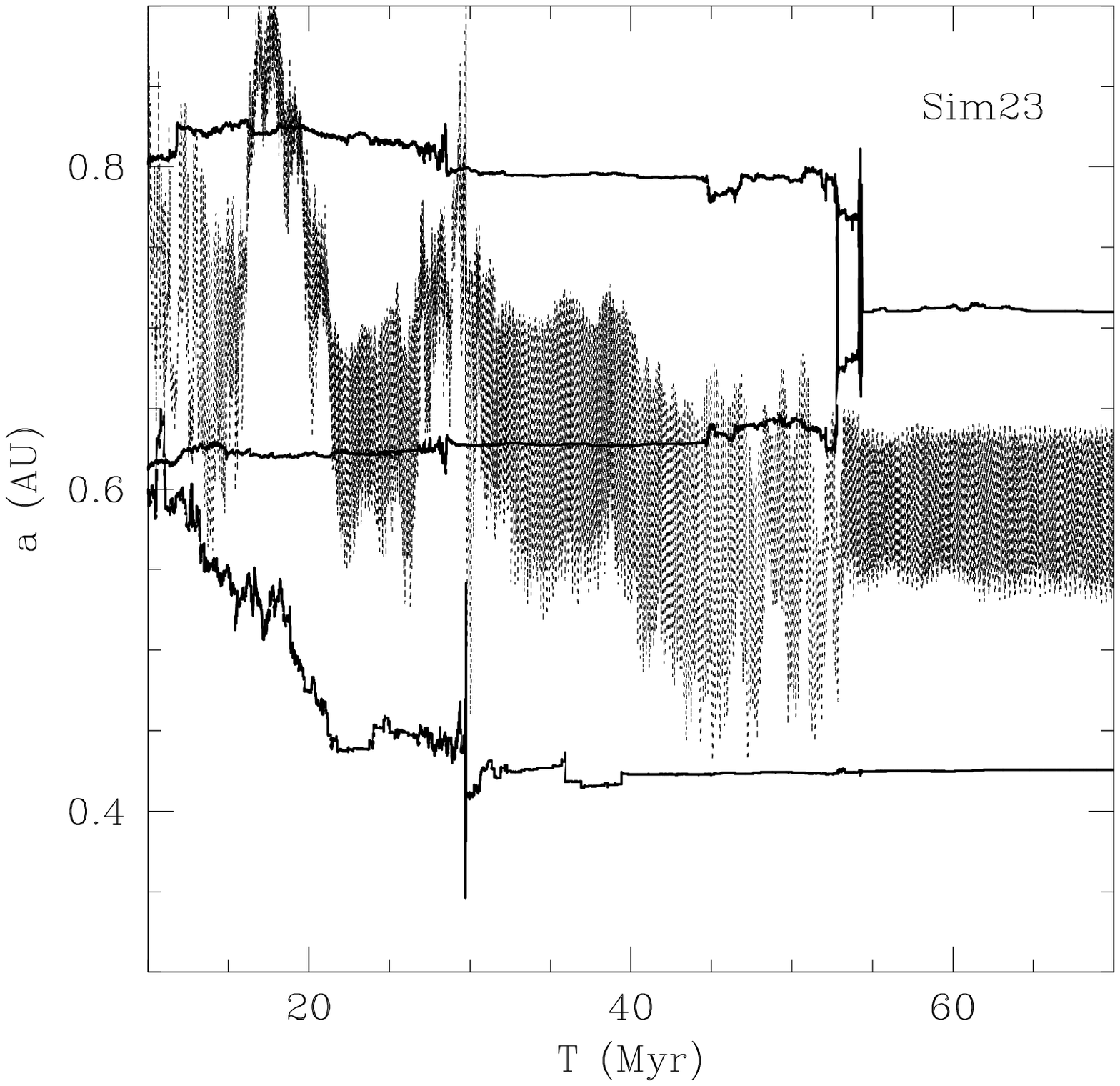}
\figcaption[Merc1]{The three solid lines show the temporal evolution of the semi-major axis for the Mercury analogue and
the two bodies which eventually combine to form the Venus analogue. The dotted line shows the apastron of the Mercury analogue.
We see that the Mercury analogue is pushed inwards by scattering off the innermost of the two Venus progenitors. Eventually,
a late impact (54 Myr) results in the merger of the two roughly equal mass bodies to form the final Venus analogue. The
collision product has a final semi-major axis intermediate between those of the progenitors, and the result is that the Mercury
analogue is no longer dynamically coupled to the outer bodies.
\label{Merc1}}

\newpage

\plotone{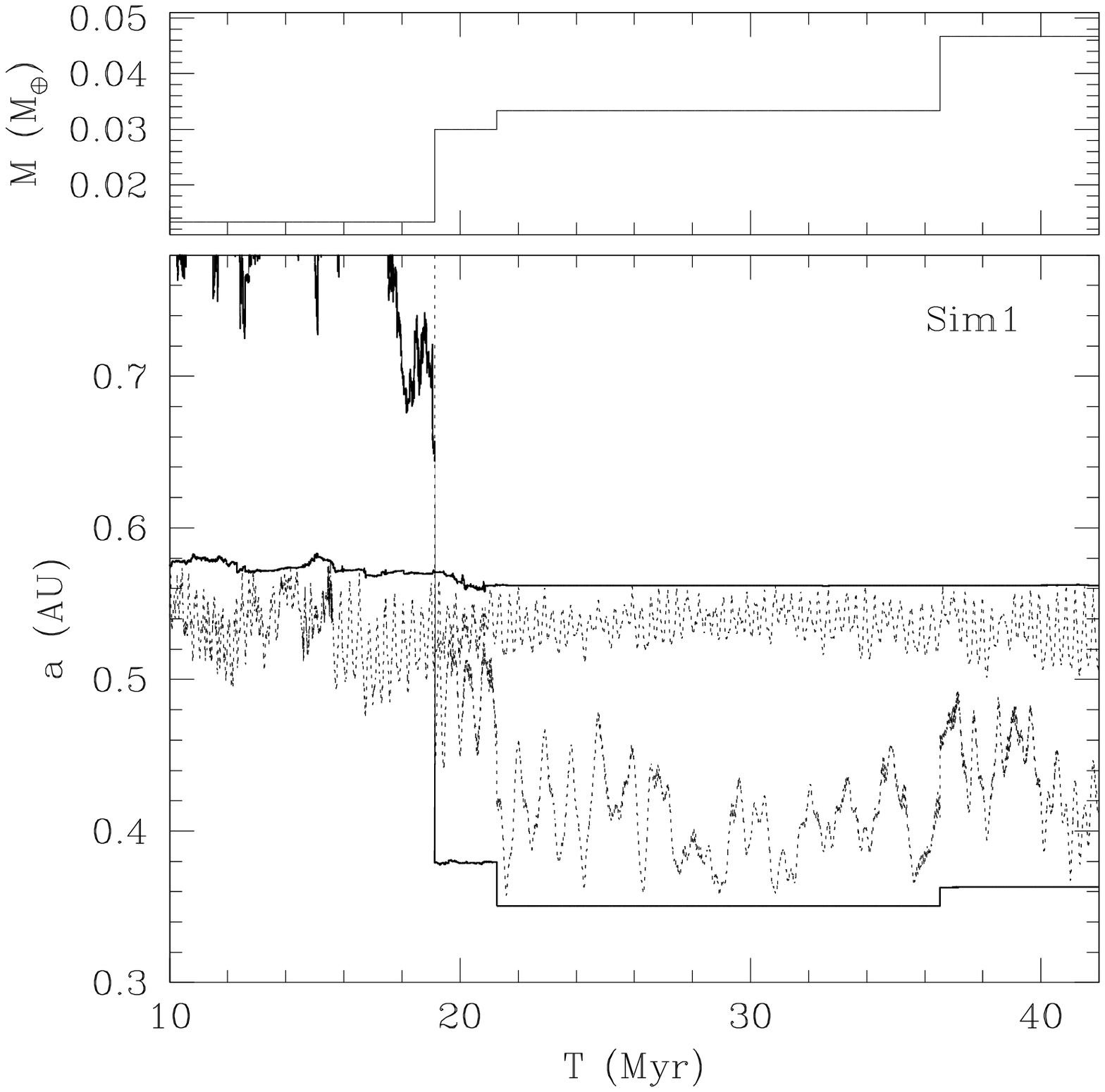}
\figcaption[Merc4]{The two solid lines in the lower panel show the evolution of the semi-major axis of the
bodies that become the Mercury and Venus analogues. The dotted lines show the evolution of the Mercury apastron
and the Venus periastron. The upper panel shows the evolution of the mass of the Mercury analogue. We see that
the decoupling from the main planet region comes as the result of two large collisions in quick succession, which
also lead to a doubling of the mass.
\label{Merc4}}

\newpage

\plotone{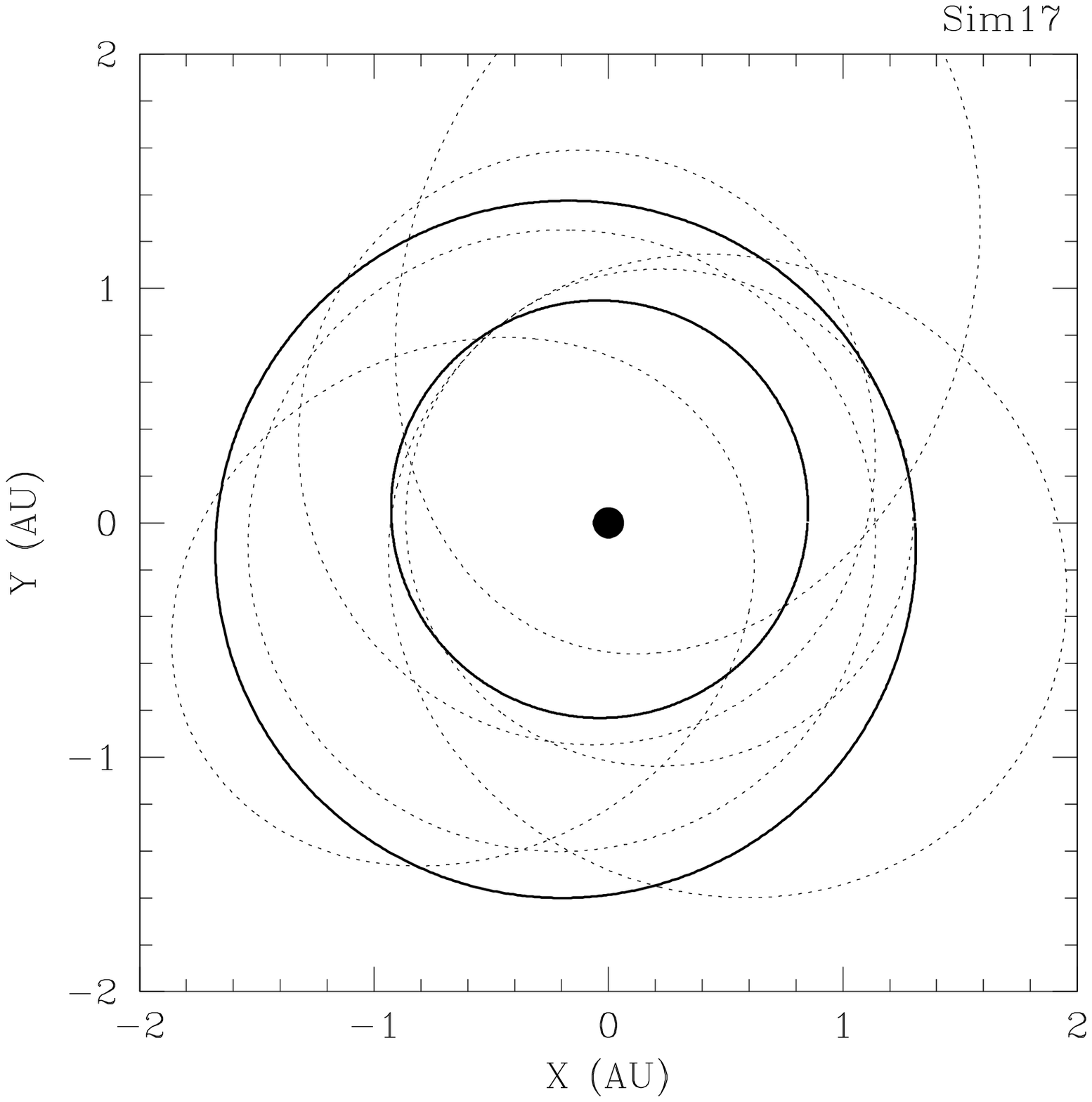}
\figcaption[Mars1]{The two solid lines show the orbits of a proto-Earth and one of the original embryos that
will eventually survive unaccreted. The snapshot is at 20 Myr. The dotted lines indicate the orbits of bodies
whose semi-major axes lie outside that of the proto-Earth at this time (all interior bodies have been omitted
for clarity). We see that the two eventual survivors are still dynamically coupled at this time but will eventually
become decoupled as the intermediaries are removed by various processes.
\label{Mars1}}

\newpage
\plotone{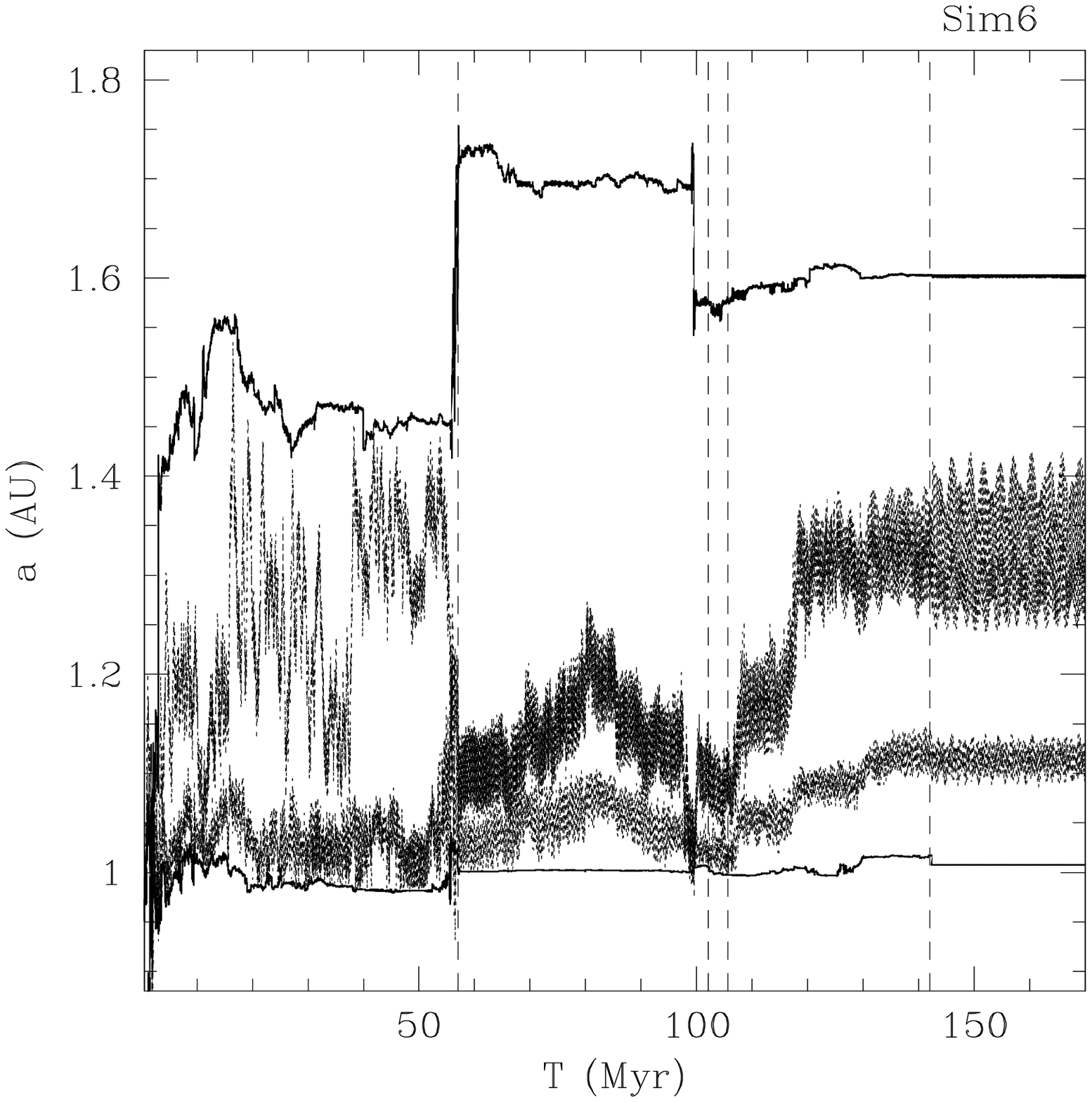}
\figcaption[Mars2]{The two solid lines show the orbital evolution of a proto-Earth and proto-Mars. The dotted
lines are the Earth's apastron and the Mars periastron. The vertical dotted lines indicate the events where
one of the intermediary bodies is removed, either by accretion or ejection.
\label{Mars2}}

\newpage

\plotone{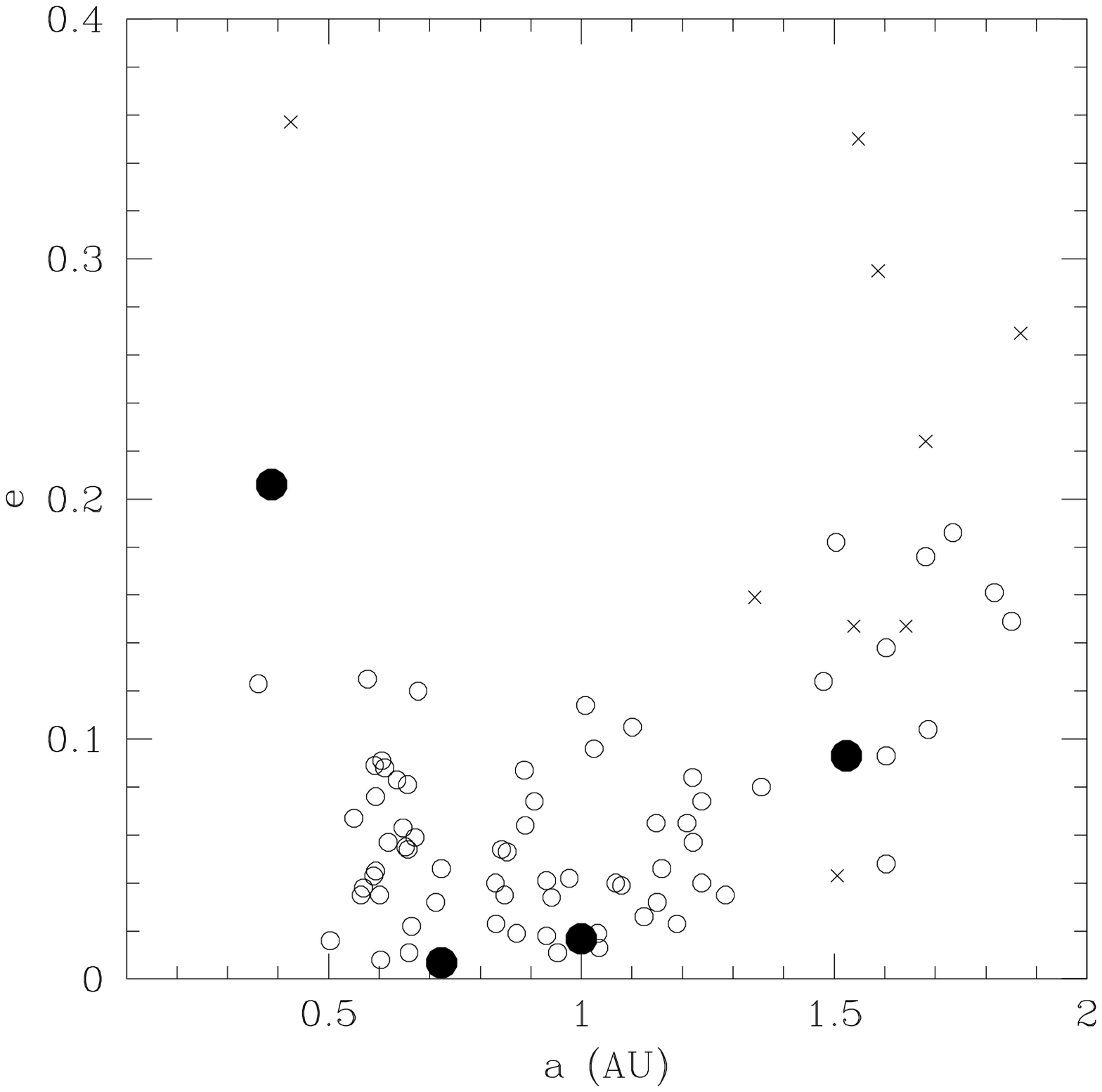}
\figcaption[ae]{The open circles are the surviving bodies (mass $> 0.02 M_{\oplus}$) in the collection
of simulations shown in Figure~\ref{am}. The bodies with mass $< 0.02 M_{\oplus}$ are shown as crosses. The four filled circles are
the parameters of the solar system planets. The increase in eccentricity as one moves away from the central annulus
is a reflection of the origin of these bodies as stranded members of diffusion tails.
\label{ae}}

\newpage

\plotone{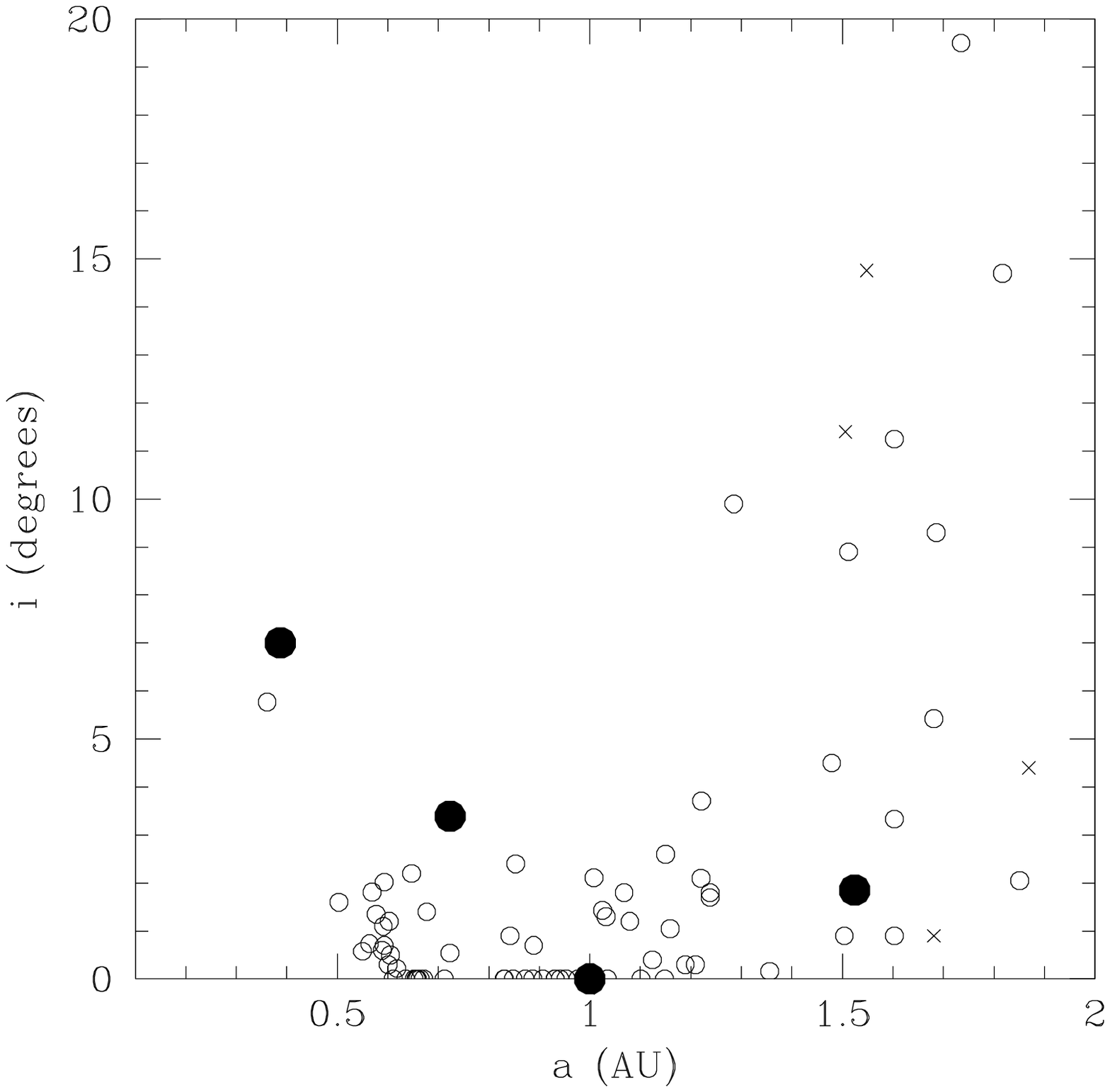}
\figcaption[ai]{The open circles show the surviving bodies with mass $>0.02 M_{\oplus}$, summed over all the simulations.
The crosses denote the remnants with masses $<0.02 M_{\oplus}$. Mercury and Mars fit well within the distributions as
does Earth (by definition, since the inclinations are expressed relative to the most massive body), although Venus seems
to have a larger than expected inclination.
\label{ai}}

\newpage

\plotone{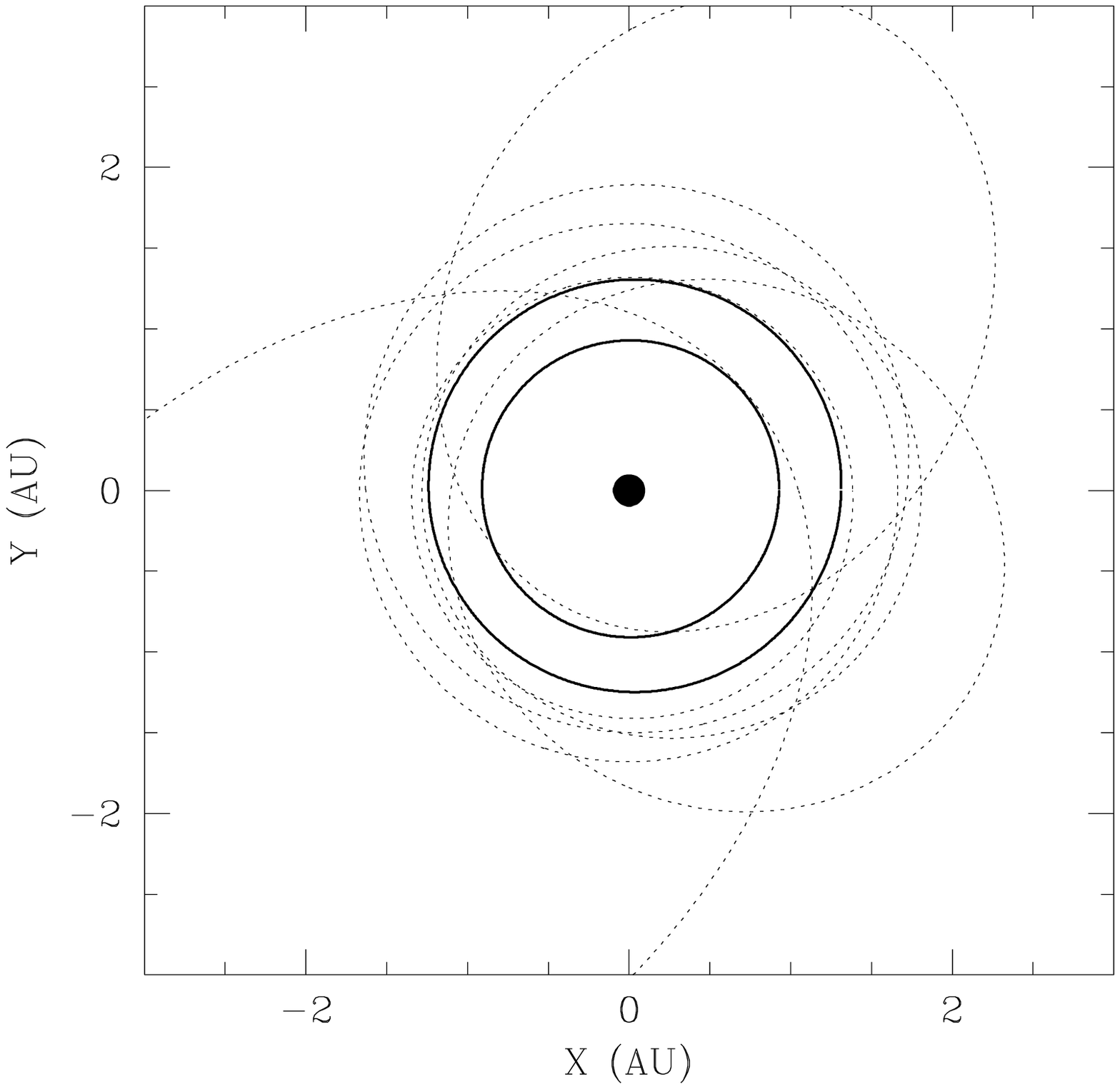}
\figcaption[NoJup]{The solid curves indicate the orbits of the bodies that will become the final Earth and Mars
analogues. The dotted lines indicate the orbits of other surviving bodies whose semi-major axes lie outside that
of the Earth analogue. This snapshot is taken after 100 Myr of evolution without any Jovian perturbations.
\label{NoJup}}

\newpage

\plotone{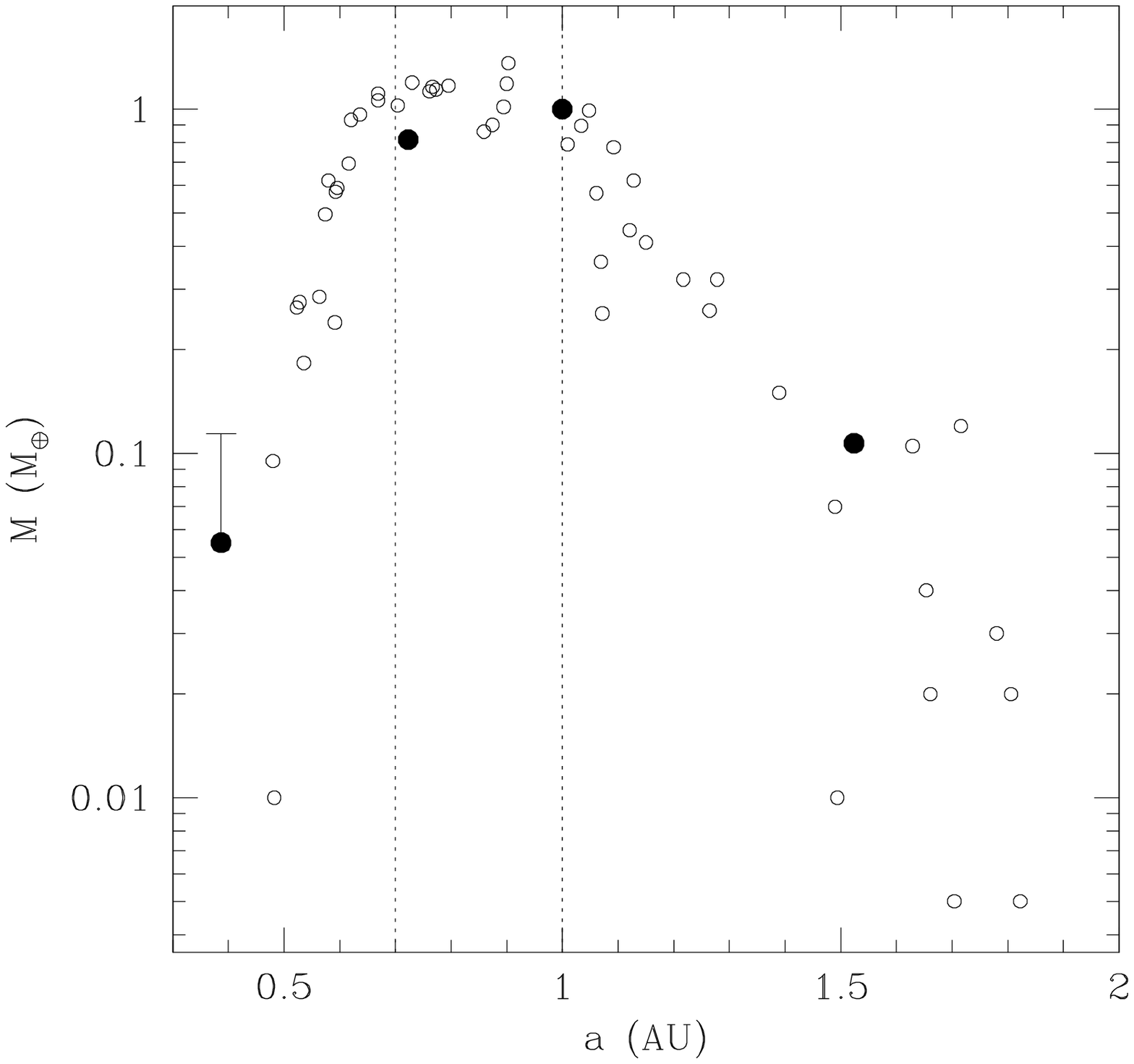}
\figcaption[am2]{The filled circles indicate the values of the known terrestrial planets, while the open circles indicate 
simulated planets after $10^9$ years in simulations where the perturbations from Jupiter were only added after 5~Myr.
\label{am2}}

\newpage
\plotone{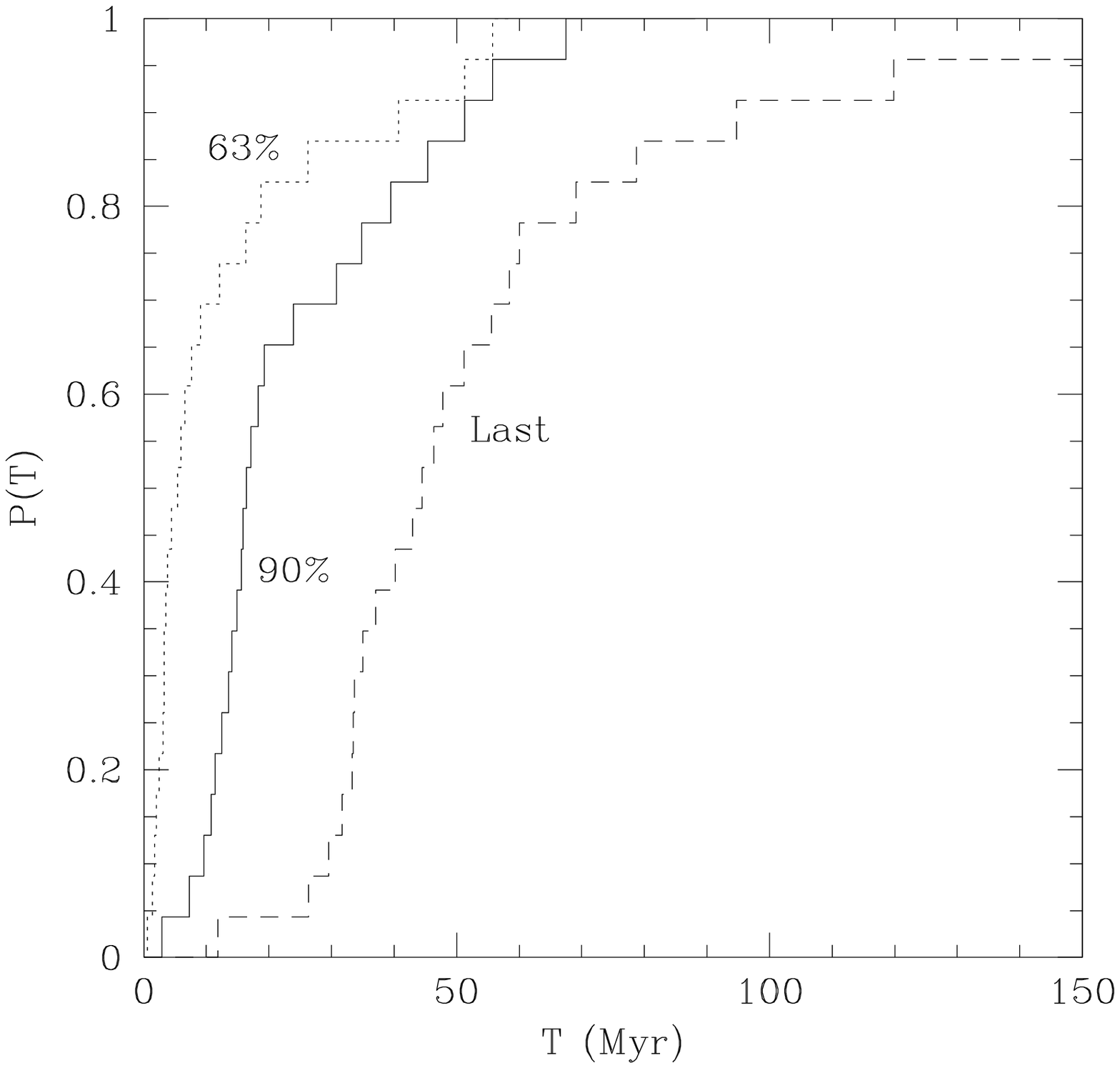}
\figcaption[Tx]{The dotted histogram shows the cumulative distribution of the time to accumulate 63\% of the mass of
the largest body in each simulation. The solid histogram shows the timescale to accumulate 90\% of the mass. The dashed
histogram indicates the distribution of the time at which the largest final body in each simulation experienced it's
last collision with a body of mass $0.02 M_{\oplus}$ or greater.
\label{Tx}}
\newpage

\plotone{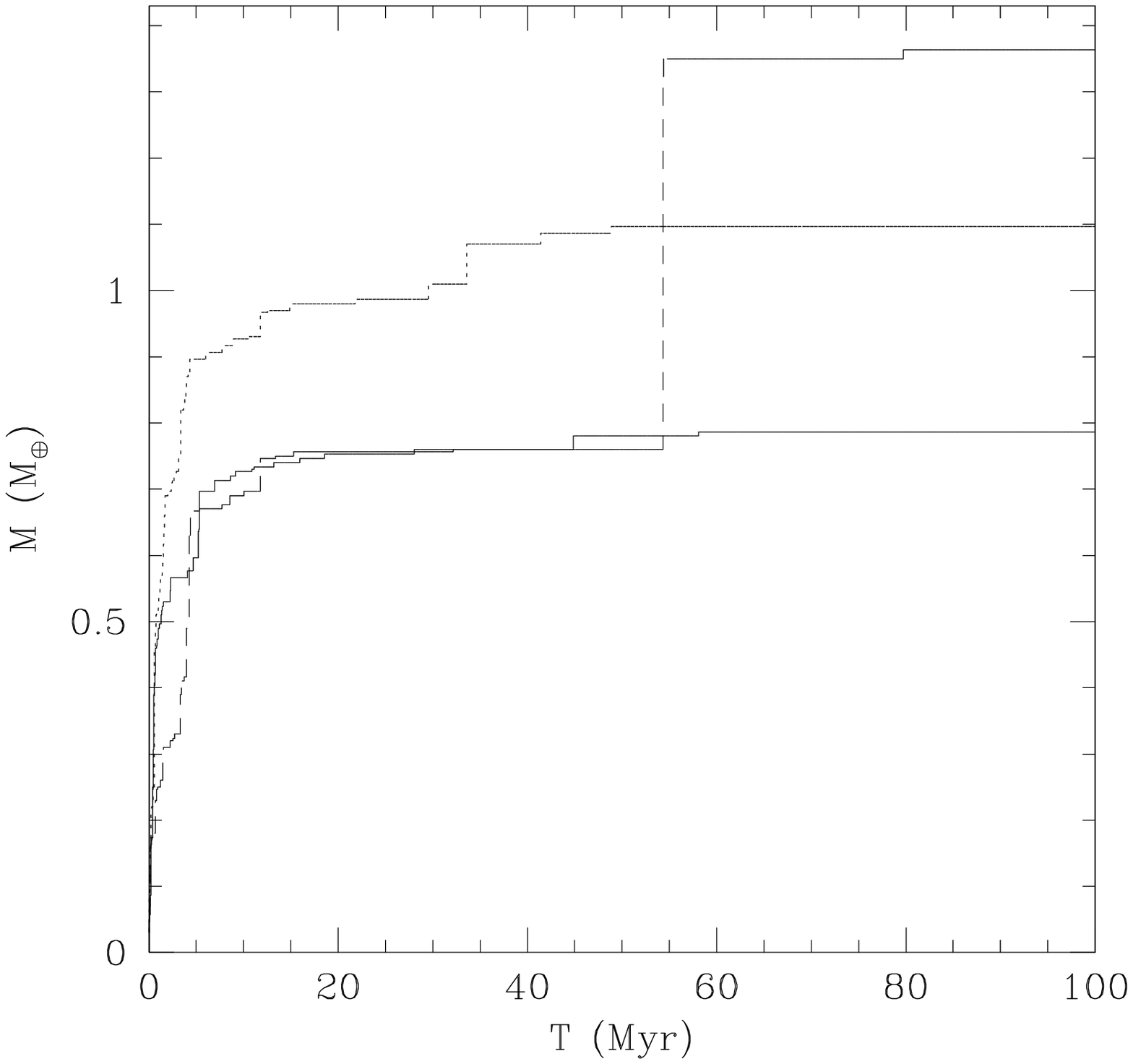}
\figcaption[Three]{The dotted, solid and dashed lines indicate the accretional history of the largest final body
in three different simulations. These show the range of potential histories, with collisions with roughly equal
mass bodies occurring both at early and late times, or not at all.
\label{Three}}

\newpage

\plotone{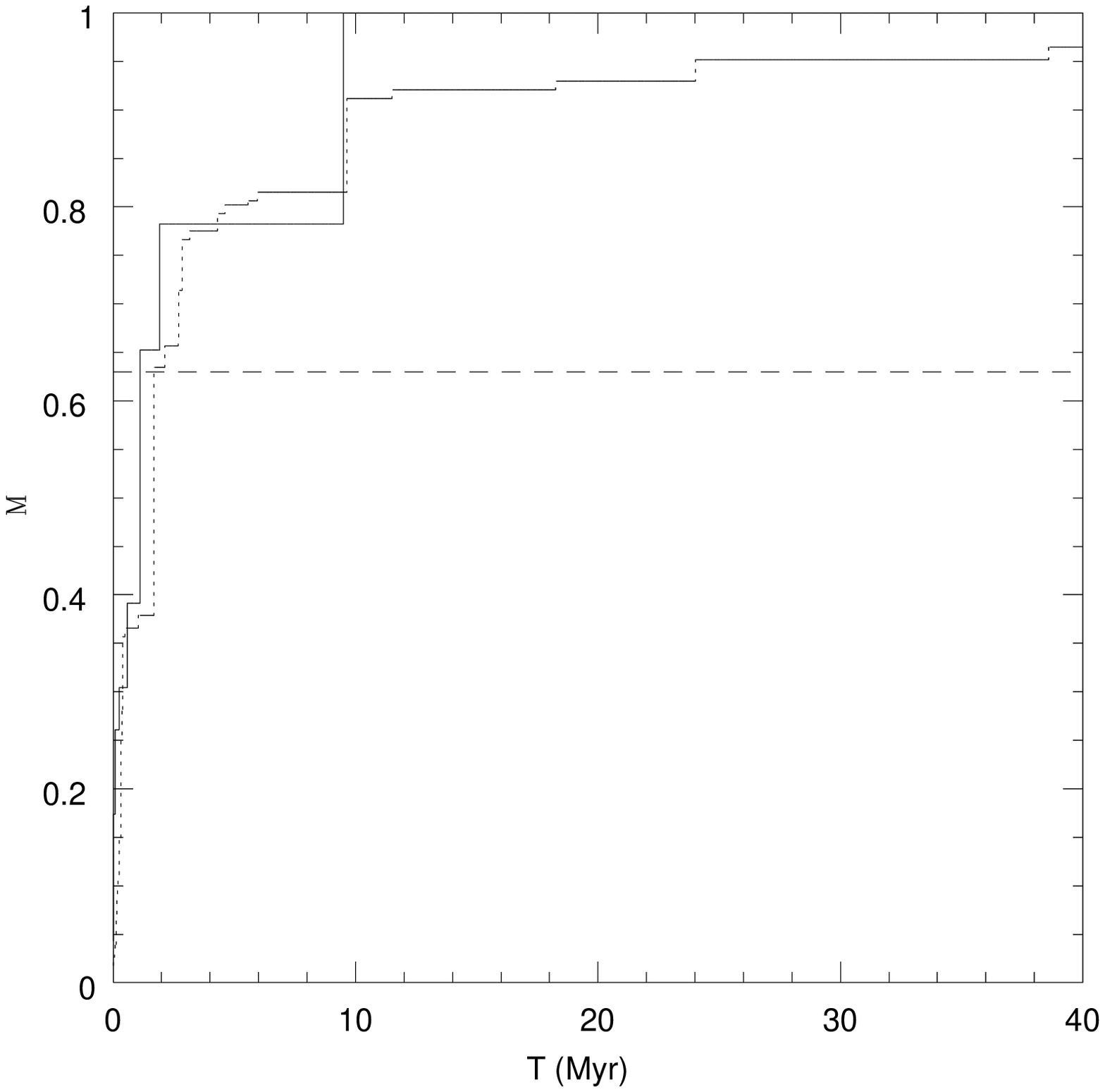}
\figcaption[TM]{The solid line indicates the accretion history of the Mars analogue in Figure~\ref{Mars2}. The dotted line
shows the history for the Earth analogue in that simulation and the dotted line indicates the nomincal 63\% threshold used
by Jacobsen (2005) to calculate the characteristic accretion time.
\label{TM}}

\newpage

\plotone{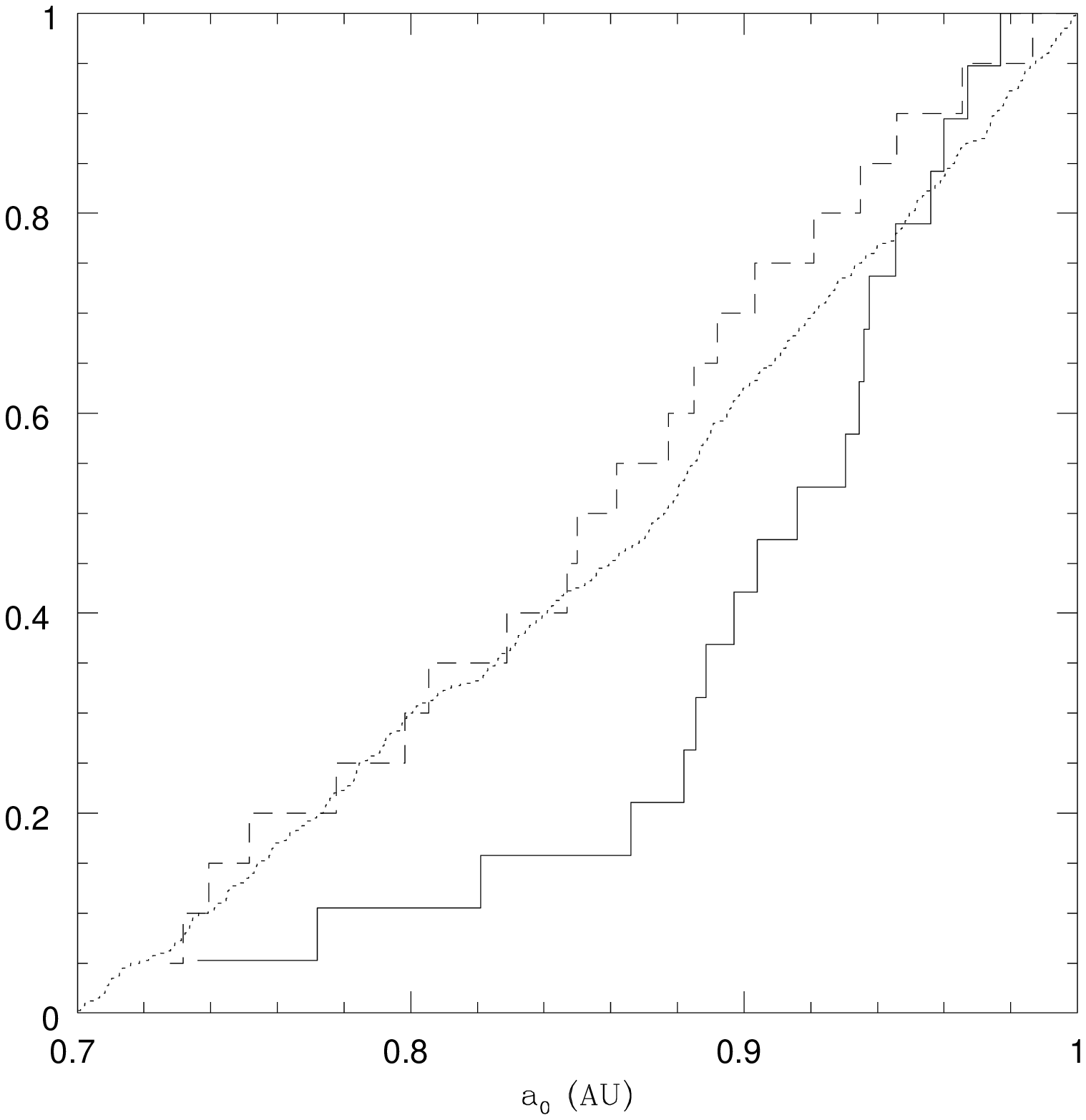}
\figcaption[Grad]{The solid and dashed histograms show the distribution of initial positions of bodies that
ended up accreting together to form the Mars analogue in two simulations (masses of $0.09 M_{\oplus}$ and
0.115$M_{\oplus}$ respectively). The dotted histogram shows the overall distribution throughout the 
annulus. We see that, in one case, the Mars is composed primarily of material from the outer part of the
nebula, while, in the other case, the material is a well-mixed representation of the entire annulus.
\label{Grad}}

\newpage

\begin{deluxetable}{lcccc} 
\tablecolumns{5} 
\tablewidth{0pc} 
\tablecaption{Planetary System Statistics for simulations of terrestrial planet assembly
 \label{CompTab}} 
\tablehead{ 
\colhead{Name}    &  \colhead{N} & \colhead{$S_d$} & \colhead{$S_s$} & \colhead{$S_c$} \\
 }
\startdata 
\sidehead{Simulations: Jupiter \@ 0 Myr}
 Sim1 & 4 & 0.00066 & 36.8 & 68.4 \\
 Sim2 & 4 & 0.00501 & 29.0 & 77.1 \\
 Sim3 & 3 & 0.00508 & 39.1 & 69.5 \\
 Sim4 & 4 & 0.00213 & 27.9 & 54.8 \\
 Sim5 & 4 & 0.00157 & 28.2 & 110.3 \\
 Sim6 & 3 & 0.00405 & 26.7 & 103.5 \\
 Sim7 & 3 & 0.00133 & 30.5 & 103.0 \\
 Sim8 & 3 & 0.00092 & 32.8 & 97.7 \\
 Sim9 & 4 & 0.00129 & 32.9 & 78.2 \\
 Sim10 & 4 & 0.00106 & 31.1 & 74.3 \\
 Sim11 & 3 & 0.00551 & 39.2 & 74.9 \\
 Sim12 & 3 & 0.00194 & 43.0 & 97.1 \\
 Sim13 & 3 & 0.00228 & 30.8 & 96.0 \\
 Sim14 & 3 & 0.00173 & 27.3 & 101.1 \\
 Sim15 & 4 & 0.00159 & 30.4 & 108.0 \\
 Sim16 & 4 & 0.00464 & 30.9 & 73.3 \\
 Sim17 & 4 & 0.00335 & 31.9 & 98.4 \\
 Sim18 & 2 & 0.00129 & 37.4 & 96.3 \\
 Sim19 & 3 & 0.00112 & 35.0 & 67.0 \\
 Sim20 & 3 & 0.00272 & 37.4 & 79.6 \\
 Sim21 & 3 & 0.00211 & 30.2 & 67.6 \\
 Sim22 & 4 & 0.00087 & 30.3 & 75.6 \\
 Sim23 & 4 & 0.00245 & 39.0 & 131.1 \\
\sidehead{Simulations: Jupiter \@5Myr}
 Sim24 & 5 & 0.00074 & 27.9 & 74.3 \\
 Sim25 & 3 & 0.00107 & 40.5 & 72.7 \\
 Sim26 & 4 & 0.00238 & 36.7 & 83.0 \\
 Sim27 & 4 & 0.00435 & 33.6 & 61.2 \\
 Sim28 & 3 & 0.00196 & 37.1 & 81.3 \\
 Sim29 & 4 & 0.00137 & 27.6 & 108.1 \\
 Sim30 & 3 & 0.00043 & 35.9 & 89.9 \\
 Sim31 & 2 & 0.00484 & 36.2 & 132.3 \\
 Sim32 & 4 & 0.00146 & 28.6 & 96.1 \\
 Sim33 & 3 & 0.00074 & 29.4 & 92.0 \\
 Sim34 & 4 & 0.00188 & 35.1 & 76.5 \\
 Sim35 & 3 & 0.00324 & 39.1 & 89.2 \\
 Sim36 & 3 & 0.00227 & 25.6 & 111.0 \\
 Sim37 & 4 & 0.00304 & 34.1 & 93.2 \\
 Sim38 & 3 & 0.00202 & 33.2 & 76.8 \\
\sidehead{Solar System}
MVEM  & 4 &  0.0018 & 37.7 & 90 \\
\enddata 
\end{deluxetable} 

\end{document}